\documentclass[%
 reprint,
 amsmath,amssymb,
 aps,
]{revtex4-2}

\usepackage{graphicx}
\usepackage{dcolumn}
\usepackage{bm}
\usepackage{hyperref}
\usepackage{float}

\begin{document}

\preprint{APS/123-QED}

\title{High-Precision Excited-State Nuclear Recoil Spectroscopy with Superconducting Sensors}

\author{C.~Bray$^{1,2}$}\email{cbray@mines.edu}
\author{S.~Fretwell$^{1,2}$}
\author{L.~A.~Zepeda-Ruiz$^2$}
\author{I.~Kim$^{2}$}
\author{A.~Samanta$^{2}$}
\author{K.~Wang$^{2,3}$}
\author{C.~Stone-Whitehead$^{1}$}
\author{W.~K.~Warburton$^{4}$}
\author{F.~Ponce$^{5}$}
\author{K.~G.~Leach$^{1,6}$}

\author{R.~Abells$^{7}$}

\author{P.~Amaro$^{8}$}

\author{A.~Andoche$^{9}$}

\author{R.~Cantor$^{10}$}

\author{D.~Diercks$^{1}$}

\author{M.~Guerra$^{8}$}

\author{A.~Hall$^{10}$}

\author{C.~Harris$^{1}$}

\author{J.~Harris$^{4}$}

\author{L.~Hayen$^{11}$}

\author{P.~A.~Hervieux$^{9}$}

\author{G.~B.~Kim$^{2}$}

\author{A.~Lennarz$^{7}$}

\author{V.~Lordi$^{2}$}

\author{J.~Machado$^{8}$}

\author{P.~Machule$^{7}$}

\author{A.~Marino$^{1}$}

\author{D.~McKeen$^{7}$}

\author{X.~Mougeot$^{12}$}

\author{C.~Ruiz$^{7}$}

\author{J.~P.~Santos$^{8}$}

\author{J.~Smolsky$^{1}$}

\author{B.~D.~Waters$^{1}$}

\author{S.~Friedrich$^{2}$}

\affiliation{$^1$Colorado School of Mines, Golden, Colorado, USA}

\affiliation{$^2$Lawrence Livermore National Laboratory, Livermore, California, USA}

\affiliation{$^3$Cornell University, Ithaca, New York, USA}

\affiliation{$^4$XIA LLC, Oakland, California, USA}

\affiliation{$^5$Pacific Northwest National Laboratory, Richland, Washington, USA}

\affiliation{$^6$Facility for Rare Isotope Beams, Michigan State University, East Lansing, Michigan, USA}

\affiliation{$^7$TRIUMF, Vancouver, BC, Canada}

\affiliation{$^8$NOVA School of Science and Technology, Lisbon, Portugal}

\affiliation{$^9$Universit{\'e} de Strasbourg, CNRS, Institut de Physique et Chimie des Mat{\'e}riaux de Strasbourg, Strasbourg, France}

\affiliation{$^{10}$STAR Cryoelectronics, Santa Fe, New Mexico, USA}

\affiliation{$^{11}$Laboratoire de Physique Corpusculaire, CNRS/IN2P3, Caen, France}

\affiliation{$^{12}$Universit{\'e} Paris-Saclay, CEA, List, Laboratoire National Henri Becquerel (LNE-LNHB), Palaiseau, France}

\collaboration{BeEST Collaboration}

\date{\today}

\begin{abstract}

Superconducting sensors doped with rare isotopes have recently demonstrated powerful sensing performance for sub-keV radiation from nuclear decay. Here, we report the first high-resolution recoil spectroscopy of a single, selected nuclear state using superconducting tunnel junction (STJ) sensors. The STJ sensors were used to measure the eV-scale nuclear recoils produced in $^7$Be electron capture decay in coincidence with a 478 keV $\gamma$-ray emitted in decays to the lowest-lying excited nuclear state in $^7$Li. Details of the Doppler broadened recoil spectrum depend on the slow-down dynamics of the recoil ion. The measured spectral broadening is compared to empirical stopping power models as well as modern molecular dynamics simulations at low energy. The results have implications in several areas from nuclear structure and stopping powers at eV-scale energies to direct searches for dark matter, neutrino mass measurements, and other physics beyond the standard model.

\end{abstract}

\maketitle

Decay spectroscopy is perhaps the most widely used experimental tool in both fundamental and applied nuclear science~\cite{PhysRevResearch.4.021001, Knoll}. In the past decade, the use of cryogenic sensors such as transition edge sensors (TESs), metallic magnetic calorimeters (MMCs), and superconducting tunnel junctions (STJs) have extended access to low-energy radiation emitted in various forms of nuclear decay~\cite{Gastaldo2017, BORGHESI2023168205, Leach2022,Fretwell2020,Friedrich2021,smolsky2024}. These sensors are characterized by their exceptionally high energy resolution and low detection thresholds -- both of which are on the order of a few eV~\cite{Bray2024}. As a result, they provide unique access to high-resolution measurements of the energy of recoiling nuclei and other sub-keV radiation that is emitted following nuclear decay. These developments have so far been driven by major focused efforts in neutrino physics using electron capture (EC) decay isotopes such as $^{163}$Ho~\cite{Gastaldo2017, BORGHESI2023168205} or $^7$Be~\cite{Leach2022,Fretwell2020,Friedrich2021,smolsky2024}. Additionally, sub-keV nuclear recoil measurements enabled by cryogenic sensors have become a focus of dark matter (DM) searches~\cite{essig2023snowmass2021,supercdmscollaboration2023strategylowmassdarkmatter,Angloher_2023,kim2024athermalphononcollectionefficiency,PhysRevD.99.123005,PhysRevD.86.051701,Carter2017} and coherent elastic neutrino-nucleus scattering (CE$\nu$NS) measurements~\cite{Rothe2020,Augier2023}. In these applications, the physics of low-energy nuclear recoils impacts the experimental sensitivity~\cite{essig2023snowmass2021,annurev-nucl-111722-025122,PhysRevD.102.063026,PhysRevLett.130.211802,PhysRevD.106.063012} and it becomes critical to understand the difference between the measured signal for a nuclear recoil (NR) and an electronic recoil (ER)~\cite{PhysRevLett.131.091801,PhysRevLett.131.091801,essig2023snowmass2021,PhysRevD.102.063026,PhysRev.124.128}. However, uncertainty surrounding ion stopping processes causes significant tension between different models and measurements of NR signals, particularly at low energy~\cite{PhysRevLett.131.091801,PhysRevD.102.063026,kavner2024measurement,PhysRevD.94.082007,PhysRev.124.128}.

In most nuclear decay processes, the prompt radiation is accompanied by the subsequent emission of $\gamma$ rays or internal conversion electrons that result from the relaxation of excited states in the daughter nucleus.  Beyond that, more exotic radioactive decay modes of excited states are possible in rare isotopes, and their daughter isotopes can also be unstable.  As a result, experimental setups that employ coincidence spectroscopy of multiple particles from the same decay enable exceptionally selective measurements by time-correlating events from a suite of different detectors~\cite{Knoll}.  This general class of nuclear decay measurement has been employed to great effect for decades. When used with cryogenic sensors, it is also expected to have increased decay selectivity enabling future neutrino mass measurement programs with ultra-low Q values into excited nuclear states \cite{PhysRevLett.127.272301,PhysRevC.107.015504}. However, the correlation between secondary particles and prompt, high-resolution measurements of an excited nuclear decay recoil has not yet been explored.  In this Letter, we report the first such coincidence spectroscopy by tagging on the $\gamma$-ray in the EC decay of $^7$Be to an excited nuclear state at 478~keV in $^7$Li.

The isotope $^7$Be is the lightest pure EC decaying nucleus, with a half-life of $T_{1/2}=53.22(6)$~days~\cite{Til02} and a decay energy of $Q_{EC}=861.963(23)$~keV~\cite{PhysRevC.109.L022501}.  A small branch of 10.44(4)\% results in the population of a short-lived excited nuclear state in $^7$Li$^*$ that de-excites via emission of a 477.603(2)~keV $\gamma$-ray~\cite{Hel00}  with a half life of $T_{1/2}=72.8(20)$~fs~\cite{Til02} and a small internal conversion coefficient (ICC) of $\alpha\approx7\times10^{-7}$~\cite{PhysRev.114.1590,Til02}. The lifetime is sufficiently short that the $\gamma$-ray is emitted while the $^7$Li$^*$ nucleus is still slowing down after the initial 11.297(9)~eV recoil from the EC decay~\cite{AME2020,Friedrich2021}. Thus, the additional recoil energy of 17.45156(15)~eV from the $\gamma$ emission can either sum with or subtract from the original recoil depending on the relative orientation of the $\gamma$ emission and $^7$Li$^*$ momentum vector. This Doppler-broadens the recoil spectrum for decay to $^7$Li$^*$, with the details of the Doppler profile containing information about the $^7$Li$^*$ kinetic energy at the time of de-excitation. The Doppler profile from decay to $^7$Li$^*$ is therefore an excellent test case for models of recoil dynamics in solids at low energies. In addition, the $^7$Be EC in STJs (BeEST) experiment currently provides the most stringent laboratory limits on sterile neutrinos in the 100-850 keV mass range~\cite{Friedrich2021}. The current sensitivity is limited by systematic uncertainties in the final-state energy spectrum that is complicated by a combination of nuclear and atomic processes at low energies~\cite{kim2024}. Selective measurements of only the $^7$Li$^*$ recoil spectrum can separate signals of different decay channels and constrain the complex spectral features in the final fit. This can either validate existing models or provide experimental input to improve modeling and thereby increase the sensitivity to new BSM physics scenarios.

\begin{figure}
    \includegraphics[width=0.95\columnwidth]{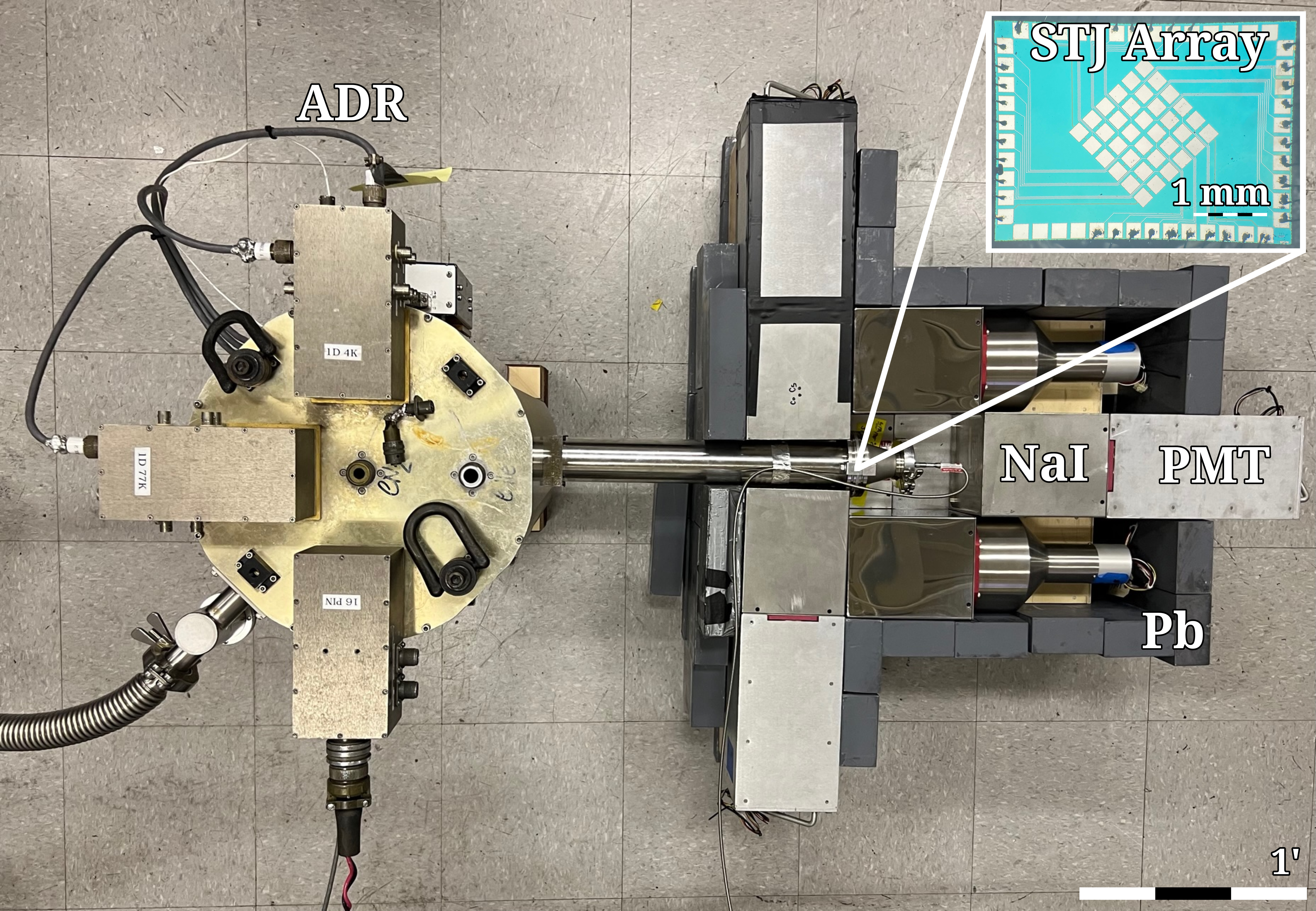}
    \caption{Photograph (top view) of the experimental setup. The 36-pixel STJ array (inset) is held at the end of the cold finger of an ADR (left) and surrounded by a NaI(Tl) array and Pb shielding (right). During operation, an additional NaI(Tl) detector and Pb shielding are added above.}
    \label{setupWithInlay}
\end{figure}

\begin{figure}[t]
    \includegraphics[width=0.95\columnwidth]{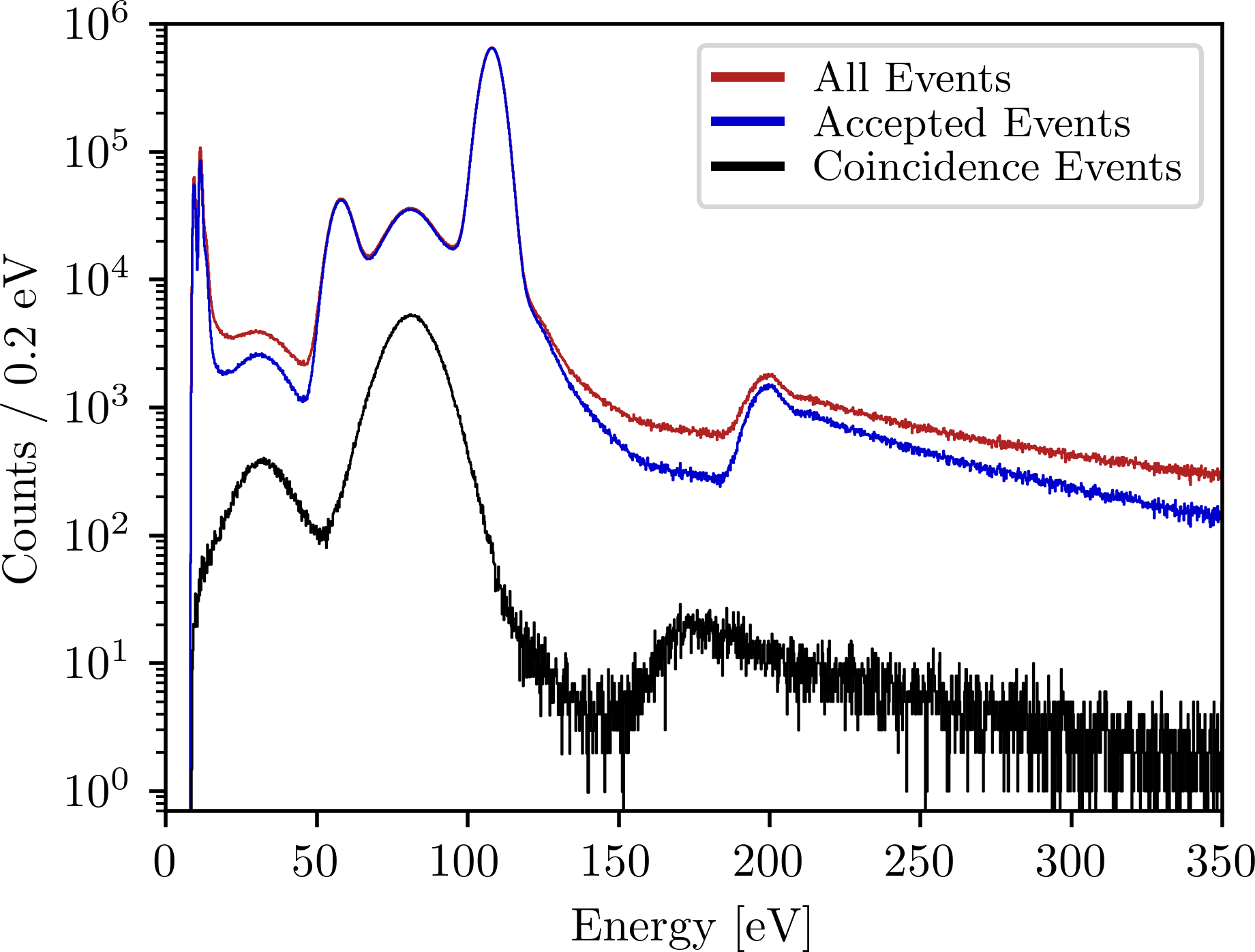}
    \caption{Summed nuclear recoil spectrum from $^7$Be EC decay measured with 14 STJ sensors over an 8-day experiment for all events (red), events that passed selection cuts (blue) and those in coincidence with a 478 keV $\gamma$-ray (black).}
    \label{fullSpectrum}
\end{figure}

The work presented here uses the same 36-pixel array of tantalum-based superconducting tunnel junction (STJ) sensors as in the earlier Phase-III of the BeEST experiment~\cite{kim2024,smolsky2024}. For that measurement, $^7$Be was implanted into the top Ta film at a voltage of 30 kV to a depth of $\sim$58 nm at TRIUMF's Isotope Separator and Accelerator (ISAC) facility~\cite{Dilling_Krucken_Merminga_2014,kim2024}. At the time of this experiment, the remaining activity was $< 20$ Bq per STJ. To achieve the required $\gamma$-tagging, we employed an adiabatic demagnetization refrigerator (ADR) that operates the sensors at the end of a cold finger $\sim$40 cm away from the main chamber of the refrigerator~(Figure \ref{setupWithInlay})~\cite{FRIEDRICH20011117}. This new experimental configuration enabled close packing of seven 5''$\times$5''$\times$6'' NaI(Tl) scintillators around the STJ sensors to capture the 478 keV $\gamma$-ray from the de-excitation of $^7$Li$^*$. The NaI(Tl) scintillators were enclosed by a layer of 2'' thick Pb bricks to reduce the ambient radioactive background count rate in the 428-528 keV region of interest by a factor of $\sim$7 to $\sim$19 counts/s total from the array. The STJ signals were amplified by a current-sensitive pre-amplifier, captured and filtered with an XIA MPX-32D digitizer, and recorded in list-mode~\cite{Warburton2015, Marino2022, Bray2024}. The $\gamma$ signals from the NaI(Tl) scintillators were processed by an XIA PixieNet-XL~\cite{pixienet} and recorded on the XIA MPX-32D when any of the NaI(Tl) detectors measured an event in the 100 keV window centered on 478 keV. This setup was used to collect 8 days of data from the 14 highest-performance STJ sensors.

\begin{figure*}[th]
    \includegraphics[width=0.95\textwidth]{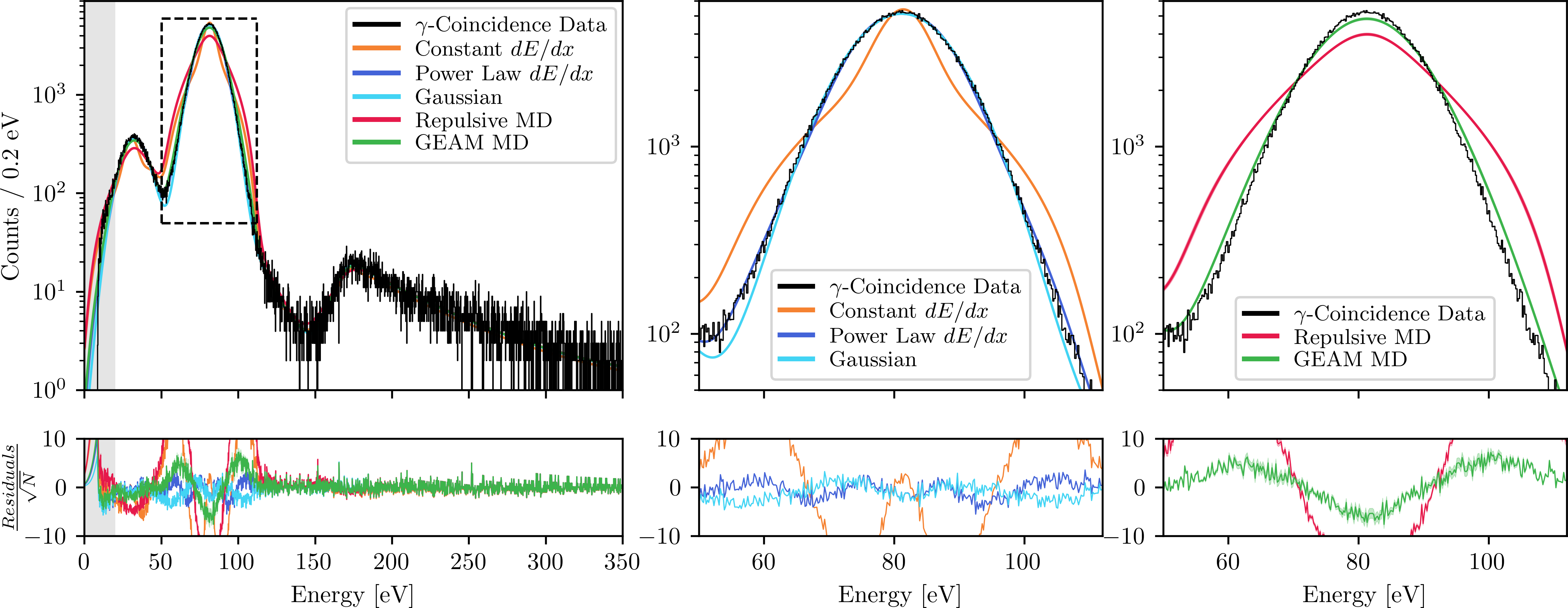}
    \caption{Comparison of the experimental data (black) with the predicted Doppler profile of the excited state recoil peaks from MD simulations, CSDA models, and Gaussian broadening. (Left) Full spectral response from models and residuals to the measured nuclear recoil spectrum. (Middle and Right) Detail region showing subtle peak shape differences between models. MD results show the average of 100 individual simulations with shaded error regions corresponding to the standard error of the mean.}
    \label{slowdownVsData}
\end{figure*}

To process this data, we first corrected the STJ signal arrival time for amplitude walk inherent to the rising edge trigger~\cite{Knoll}. Next, we vetoed any three or more coincident STJ signals within a 15 $\mu$s window to remove the broad spectral background from $\gamma$ interactions in the substrate below the STJs~\cite{kim2024}. We also rejected events when two STJs triggered within that window to avoid ambiguous $\gamma$-tagging. We then cut signals from any period where the count rate was more than 5$\sigma$ above the mean event rate to reject occasional periods of high electromagnetic pickup. We also rejected spectra whose noise peak extends above the 20 eV lower bound of the analysis range or whose detector resolution significantly broadens the $^7$Be spectrum. Finally, we drift corrected the data using a 5-minute median filter on the most prominent $^7$Be decay peak at 108.59(6) eV~\cite{kim2024}. Following the initial data acceptance cuts, we implemented a coincidence condition to select only events that result from $^7$Be decay to the excited state of $^7$Li. Specifically, we calculate the energy-dependent timing jitter for each channel and select only events within a 3$\sigma$ coincidence window around a $\gamma$ signal from NaI(Tl) between 428 and 528 keV. The sums of the spectra from 14 STJs over 8 days are shown in Figure~\ref{fullSpectrum}. 

The $\gamma$-tagged spectrum in Figure~\ref{fullSpectrum} has reduced statistics compared to the full event spectrum due to the tagging efficiency of 15.6(2)\%, but it significantly improves the ability to resolve details of the excited state decay spectrum through its high selectivity. This spectrum reveals for the first time the detailed $^7$Li$^*$ recoil profile from K-shell electron capture at 50-150 eV and L-shell electron capture at 20-50 eV, as well as the shake-off events following decay into the excited state at $>$150 eV \cite{kim2024}.

Historically, the Continuous Slowing Down Approximation (CSDA)~\cite{Carron_2006} has been used to model ion stopping profiles, assuming a constant $dE/dx$ at low energy~\cite{PhysRev.124.128,osti_4701226}~(Figure~\ref{slowdownVsData}, center). To fit the data, we use \texttt{iminuit}~\cite{iminuit,James:1975dr} to simultaneously fit the $\gamma$-coincidence and full energy spectra from each sensor, allowing variation between sensors only for the activity, resolution, noise peak, and Auger escape probability~\cite{Friedrich2021,kim2024}. The excited-state decay peaks were constrained to the modeled Doppler profile convolved with the ground-state response. In order to estimate the systematic error from the energy calibration uncertainty, the excited state lifetime uncertainty, and from fit convergence, we repeated each fit 100 times with slightly modified calibrations and $^7$Li$^*$ lifetimes. Here, we report the standard deviation of the fits as the systematic error. The best fit value for a constant $dE/dx$ is $-18.57 \pm 0.02_{stat.} \pm 0.61_{sys.}$ eV/nm with a reduced $\chi^2_\nu$ of 25.3. The modeled spectral shape clearly deviates from the data (Figure \ref{slowdownVsData}, middle), with a central peak of events that have completely stopped and a broader spectrum of events that have barely slowed. Choosing a $dE/dx$ = $a E^b$~\cite{PhysRevLett.66.1831,Sigmund2014,NORTHCLIFFE1970233} yields a considerably better profile, with a best fit for a = $-1.450 \pm 0.001_{stat.} \pm 0.044_{sys.} \textnormal{eV}^{1-b}/\textnormal{nm}$ and b = $1.4350 \pm 0.0006_{stat.} \pm 0.0030_{sys.}$ with a reduced $\chi^2_\nu$ = 2.04. 

Despite this agreement, we emphasize that we do not directly measure the stopping power as a function of energy, and while the data show a clear preference for the power law compared to the constant $dE/dx$, this does not exclude other stopping power models. For example, in other BeEST measurements, a Gaussian broadening model has been used to fit the excited-state decay peaks~\cite{kim2024,Friedrich2021}. In this dataset, we also find that Gaussian broadening describes the data well, with the best fit for $\sigma = 7.485 \pm 0.001_{stat.} \pm 0.010_{sys.}$ eV with a reduced $\chi^2_\nu$ of 2.10. This fits the data approximately as well as the power law $dE/dx$, which is surprising given the lack of physical motivation. Specifically, this model would fail to reproduce the spectral response from a material that stopped the $^7$Li$^*$ ion more slowly, such as HgTe~\cite{PhysRevLett.88.012501}. Nonetheless, the comparable $\chi^2_\nu$ value indicates that this model is sufficient for the ongoing analysis of the Phase-III BeEST data, particularly because the largest deviations from the $\gamma$-coincidence data are in the 50-60 eV and 100-110 eV regions where the spectrum is dominated by decay into the ground state.

\begin{figure}[h]
    \vspace{1em}
    \includegraphics[width=0.95\columnwidth]{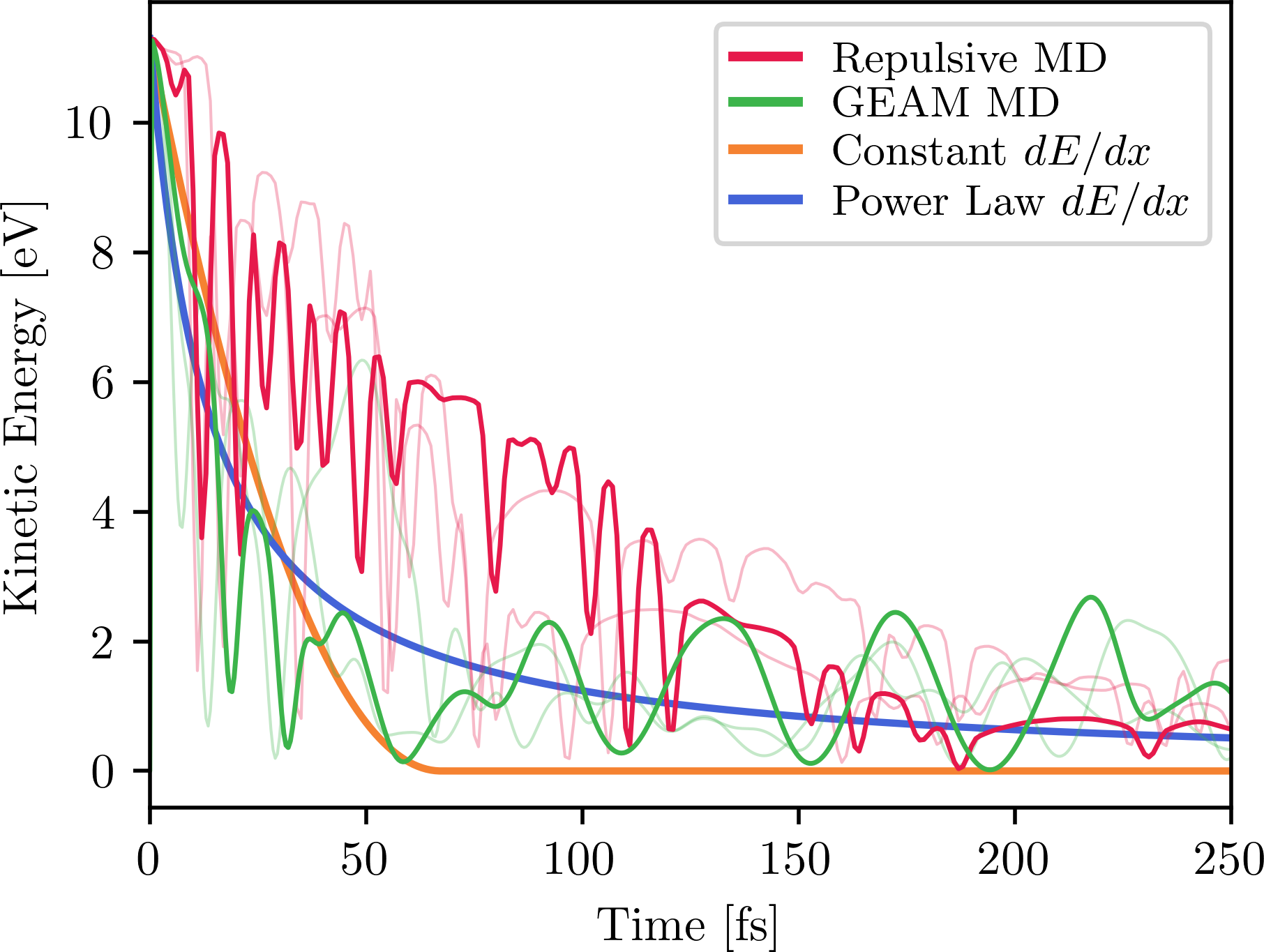}
    \caption{Predicted $^7$Li slowdown from MD simulations after EC decay imparts 11.3~eV at $t=0$. The kinetic energy evolution from 2 out of 100 MD simulations with different initial conditions are shown in transparent lines, with a typical case highlighted for both potentials. The best fit slowdown curves from the constant and power law CSDA approximations are shown for comparison.}
    \label{slowdownCurves}
\end{figure}

To better understand the Doppler profile and provide physical basis for our modeling, we have performed molecular dynamics (MD) simulations of the slow-down dynamics of $^7$Li recoils in a Ta matrix using the LAMMPS framework~\cite{lammps}. The $^7$Li$^*$ atom was initially placed in an interstitial site and given 11.3 eV of kinetic energy in a random direction. The Ta matrix was modeled using an embedded atom method (EAM) potential~\cite{eam-ta} for matrix interactions, with a repulsive pair potential initially describing Li-Ta interactions~\cite{juslin}. These simulations, run in the NPH ensemble at 0.1 K, showed recoil deceleration over approximately 100 fs, with striking dips in velocity when the Li scatters off of the Ta atoms in the matrix (Figure \ref{slowdownCurves}). However, that slowdown was approximately half the rate observed experimentally, which is far outside the measured $^7$Li$^*$ lifetime uncertainty of $\sim$3\%~\cite{Til02}. This motivated refinement using an existing generalized embedded atom method (GEAM) potential~\cite{SHI2024112732,Wang_Lordi_Friedrich_Samanta}, which incorporates two- and three-body interactions and is validated against density functional theory (DFT) calculations~\cite{PhysRevApplied.19.014032}. The Doppler profile generated by these MD results (Figure \ref{slowdownVsData}, right) revealed that the GEAM potential better reproduced the Doppler profiles observed experimentally, particularly in the 50–150 eV energy range, and highlights the sensitivity of the Doppler profile to accurate interaction modeling.

In summary, we report the first high-resolution nuclear recoil measurement of a specific excited nuclear state with superconducting sensors through coincidence spectroscopy. The selectivity of the measurement revealed the full Doppler profile and allowed us to evaluate ion slowing at eV-scale energies for the first time. The measured Doppler profile was compared with traditional CSDA models and more detailed MD simulations to demonstrate that the measurement is a sensitive test of the recoil physics. Importantly, the profile is sensitive to the specific interaction potential used in MD models of ion slow-down, enabling experimental tests of simulations with direct implications for CE$\nu$NS and DM searches~\cite{PhysRevD.106.063012}. Understanding these slow-down details will also increase the sensitivity of the BSM physics goals of the BeEST experiment by constraining key fit parameters and improving understanding of how the Doppler profile changes as a function of initial recoil energy. Further, the success of the $\gamma$-nuclear recoil coincidence technique is the first step towards cryogenic neutrino mass measurements using decays into excited states~\cite{PhysRevLett.127.272301,PhysRevC.107.015504}. Finally, this class of measurements can be used as a powerful technique for characterizing nuclear recoils in low-threshold detectors with much higher statistics than are achievable with neutron scattering or other direct reaction techniques.

\begin{acknowledgments}
The BeEST experiment is funded in part by the Gordon and Betty Moore Foundation (10.37807/GBMF11571), the DOE-SC Office of Nuclear Physics under Award Numbers DE-SC0021245, DE-FG02-93ER40789, and SCW1758, and the LLNL Laboratory Directed Research and Development (LDRD) program through Grants No. 19-FS-027 and No. 20-LW-006. TRIUMF receives federal funding via a contribution agreement with the National Research Council of Canada. The theoretical work was performed as part of the European Metrology Programme for Innovation and Research (EMPIR) Projects No. 17FUN02 MetroMMC and No. 20FUN09 PrimA-LTD. This work was performed under the auspices of the U.S. Department of Energy by Lawrence Livermore National Laboratory under Contract No. DE-AC52- 07NA27344. Francisco Ponce is funded as part of the Open Call Initiative at PNNL and conducted under the LDRD Program. PNNL is a multiprogram national laboratory operated by Battelle for the U.S. Department of Energy. This research was funded in part by FCT (Portugal) under the research center grant UID/FIS/04559/2020 (LIBPhys).
\end{acknowledgments}

\bibliography{apssamp}

\end{document}